\documentclass[prb,twocolumn]{revtex4}
\usepackage{graphicx}
\usepackage{dcolumn}
\usepackage{bm}
\usepackage{amsmath}
\usepackage{indentfirst}
\usepackage{times}
\usepackage{hhline}

\newcommand{\ph}[1]{\phantom{#1}}

\begin{document}

\title{Predicting polarization enhancement in multicomponent
ferroelectric superlattices}

\author{S. M. Nakhmanson}
\author{K. M. Rabe}
\author{David Vanderbilt}

\affiliation{Department of Physics and Astronomy, Rutgers
University, Piscataway, NJ 08854-8019}

\date{\today}

\begin{abstract}
\textit{Ab initio} calculations are utilized as an input to develop 
a simple model of polarization in epitaxial short-period 
CaTiO$_3$/SrTiO$_3$/BaTiO$_3$ superlattices grown on a SrTiO$_3$ substrate. 
The model is then combined with a genetic algorithm technique to optimize 
the arrangement of individual CaTiO$_3$, SrTiO$_3$ and BaTiO$_3$ layers
in a superlattice, predicting structures with the highest possible 
polarization and a low in-plane lattice constant mismatch with the 
substrate. This modelling procedure can be applied to a wide range
of layered perovskite-oxide nano\-structures providing guidance
for experimental development of nanoelectromechanical devices with 
substantially improved polar properties.
\end{abstract}


\maketitle

Modern epitaxial thin film techniques make it possible to synthesize
artificial multicomponent perovskite oxide superlattices whose
polar properties can be precisely tailored for a wide variety of 
applications.~\cite{yamada2002,eckstein2003,lee2005} For example, it 
was recently demonstrated that hundreds of atomically thin individual layers of
CaTiO$_3$ (CT), SrTiO$_3$ (ST) and BaTiO$_3$ (BT) could be grown on
a pe\-rov\-skite ST substrate, yielding superlattices with compositionally 
abrupt interfaces and atomically smooth surfaces.~\cite{eckstein2003,lee2005}
It was also shown that --- since relaxed lattice constants of CT and BT are 
0.07 \AA\ smaller and 0.11 \AA\ larger than that of ST ($a_\mathrm{ST} = $3.905 \AA), 
respectively --- epitaxial strain in the constituent layers of these
structures can be substantial. Due to the strong coupling between
strain and polarization in ferroelectric perovskites, this can result 
in substantial enhancement of the polarization relative to that of the bulk constituents,
as has been observed~\cite{lee2005} in accordance with theoretical 
predictions~\cite{neaton2003}. 

We recently studied such strain-induced polarization enhancement in two- 
and three-component ferroelectric CT/ST/BT superlattices 
epitaxially matched to a cubic ST substrate.~\cite{serge_APL} First 
principles methods, namely density-functional theory and the modern theory
of polarization,~\cite{mpt} were used to compute the structure and
polarization of a small number of 
short-period structures with the same or similar compositions as those 
grown and characterized by Lee \textit{et al.}~\cite{lee2005,ho_nyung}
Unfortunately, the substantial computational costs associated with
these first-principles techniques, growing rapidly as the period of the
superlattice increases, make it impossible to perform the calculations
necessary to answer a broader
and more interesting question: how should we arrange individual CT, ST 
and BT layers in a given superlattice to obtain the largest possible polarization 
enhancement? In this paper we address this question by using our 
\textit{ab initio} results~\cite{serge_APL} as an input to create a simple model 
for polarization in CT/BT/ST superlattices, and then employing this model in 
conjunction with a genetic optimization algorithm technique to identify the 
optimal candidate structures.

In previous work,\cite{neaton2003} a simple continuum model 
was introduced based on first-principles calculations of ST/BT superlattices;
a similar model was subsequently applied to ST/PT superlattices.\cite{karen2005}
The main premise was to assume that the constituent layers were linear dielectrics
(in the case of the ferroelectric constituent, possessing also a nonzero
spontaneous polarization), and to obtain the value of the uniform polarization in each
layer by solving the equations of macroscopic electrostatics. With appropriate
choices for the two dielectric constants, this model could
reproduce the approximate constancy
of the local polarization in the superlattice, giving a nonzero polarization in 
the ST layer, and the dependence of the 
polarization on the ratio of the thickness of the ST and BT layers.
However, the electrostatic continuum character of this model could not reproduce the
dependence of the polarization on the absolute thickness of the
constituent layers, clearly present in our first-principles
results for CT/ST/BT superlattices. For example, the polarization of the
(ST)$_1$(BT)$_1$ superlattice is noticeably smaller than that of (ST)$_2$(BT)$_2$.

Here, we introduce a model that includes this ``size effect," based on
the following expression for the energy of the superlattice as a function 
of the scalar polarization $p_i$ of 
individual unit-cell layers $i$:
\begin{equation}
E=\sum_i (\alpha_i p_i^2 + \beta_i p_i^4) + \sum_i J_{i,i+1}\,p_i\,p_{i+1}
\;.
\label{dva}
\end{equation}
Here $\alpha_i$ and $\beta_i$ describe the anharmonic
potential of a single unit-cell layer, and $J_{i,i+1}$ represents the 
coupling between nearest-neighbor layers.  These parameters take values
that depend on the identity of the layer; for example,
$\alpha_i$ takes the values $\alpha_C,\;\alpha_S,\;\alpha_B$
for a CT, ST, or BT layer respectively. Similarly,
$\beta_i$ takes the values $\beta_C,\;\beta_S,\;\beta_B$,
and there are six interface terms
$J_{CC}$, $J_{BB}$, $J_{SS}$, $J_{SC}$, $J_{SB}$, 
and $J_{BC}$.

\begingroup
\squeezetable
\begin{table*}
\tabcolsep=0.15cm
\renewcommand{\arraystretch}{1.3}

\caption{Columns two and three contain expressions for quadratic ($A$) and quatric ($B$) energy-decomposition
coefficients for the two- and three-component superlattices of Ref.~\onlinecite{serge_APL}.
The \textit{ab initio} ground-state superlattice energies relative to the nonpolar structures, the values of 
\textit{ab initio} and fitted superlattice polarizations, as well as the differences between
the two, are shown in columns four, five, six and seven respectively.
}
\label{AB_coeff}
\begin{ruledtabular}
\begin{tabular}{lcccccc}

  System    & $A$ & $B$ & $\Delta E$ (eV) & $|p_\mathrm{min}|$ (C/m$^2$) &
  $|p_\mathrm{min}^{\mathrm{fit}}|$ (C/m$^2$) & $|\Delta p_\mathrm{min}|$ (\%) \\
 		       		       
\hline		       		       
		       		       
  Strained bulk:   \\

  CT     &  $\widetilde{J}_{CC}$ &  $\beta_C$  & -0.019 & 0.434 & 0.370 & 14.85  \\

  BT     &  $\widetilde{J}_{BB}$ &  $\beta_B$  & -0.044 & 0.368 & 0.363 & \ph{1}1.30   \\

\hline

  Two-component:   \\

  (CT)$_1$(ST)$_1$  &  $2\widetilde{J}_{CS}$ &  $\beta_C+\beta_S$ & -0.005 & 0.026 & 0.026 & \\

  (ST)$_1$(BT)$_1$  &  $2\widetilde{J}_{SB}$ &  $\beta_S+\beta_B$ & -0.025 & 0.231 & 0.231 & \\

  (CT)$_1$(BT)$_1$  &  $2\widetilde{J}_{CB}$ &  $\beta_C+\beta_B$ & -0.039 & 0.231 & 0.231 & \\

  (CT)$_2$(ST)$_2$  &  $\widetilde{J}_{SS}+\widetilde{J}_{CC}+2\widetilde{J}_{CS}$ &  
                       $2\beta_C+2\beta_S$   & -0.007 &  0.168  &  0.168 &  \\

  (ST)$_2$(BT)$_2$  &  $\widetilde{J}_{SS}+\widetilde{J}_{BB}+2\widetilde{J}_{SB}$ &  
                       $2\beta_S+2\beta_B$   & -0.039 &  0.245  &  0.245 &  \\

  (CT)$_2$(BT)$_2$  &  $\widetilde{J}_{CC}+\widetilde{J}_{BB}+2\widetilde{J}_{CB}$ &  
                       $2\beta_C+2\beta_B$   & -0.081 &  0.306  &  0.306 &  \\

\hline

  Three-component:   \\					    				    

  (CT)$_1$(ST)$_1$(BT)$_1$  &  $\widetilde{J}_{CS}+\widetilde{J}_{SB}+\widetilde{J}_{CB}$ &
                               $\beta_C+\beta_S+\beta_B$ & -0.034 & 0.200 & 0.194 & \ph{1}2.98 \\

  (CT)$_2$(ST)$_2$(BT)$_2$  &  $\widetilde{J}_{CC}+\widetilde{J}_{SS}+\widetilde{J}_{BB}+
                                \widetilde{J}_{CS}+\widetilde{J}_{SB}+\widetilde{J}_{CB}$ &
                               $2\beta_C+2\beta_S+2\beta_B$ & -0.057 & 0.242 & 0.249 & \ph{1}2.94  \\

  (CT)$_2$(ST)$_2$(BT)$_4$  &  $\widetilde{J}_{CC}+\widetilde{J}_{SS}+3\widetilde{J}_{BB}+
                                \widetilde{J}_{CS}+\widetilde{J}_{SB}+\widetilde{J}_{CB}$ &  
                               $2\beta_C+2\beta_S+4\beta_B$ & -0.131 & 0.298 & 0.287 & \ph{1}3.83 \\  

  (CT)$_3$(ST)$_3$(BT)$_3$  &  $2\widetilde{J}_{CC}+2\widetilde{J}_{SS}+2\widetilde{J}_{BB}+
                                \widetilde{J}_{CS}+\widetilde{J}_{SB}+\widetilde{J}_{CB}$ & 
                               $3\beta_C+3\beta_S+3\beta_B$ & -0.082 & 0.260 & 0.265 & \ph{1}1.91    \\

\end{tabular}
\end{ruledtabular}
\end{table*}
\endgroup

To compute the energy for arbitrary values of the unit-cell layer
polarizations $p_i$ would thus require knowledge of twelve parameters.
However, the approximate constancy of the polarization across unit-cell
layers observed in the first-principles results suggests a simplification in
which, for each superlattice, $p_i$ is taken to be uniform and equal to the 
overall polarization.
Substituting $p_i=p$
into Eq.~(\ref{dva}), we find that
\begin{equation}
E(p) = A p^2 + B p^4
\label{eq2}
\end{equation}
where
\begin{equation}
A=\sum_\nu N_\nu\alpha_\nu + \sum_{\langle\nu\nu'\rangle}
   N_{\nu\nu'} J_{\nu\nu'} \;,
\label{dvb}
\end{equation}
\begin{equation}
B=\sum_\nu N_\nu\beta_\nu \;,
\label{dvc}
\end{equation}
and $N_\nu$ and $N_{\nu\nu'}$ are the number of
layers of type $\nu$ and the number of interfaces of type
${\nu\nu'}$ appearing in the superlattice sequence.
The fact that $N_C=N_{CC}+(N_{CS}+N_{CB})/2$, and similarly for $N_S$ and 
$N_B$, for any periodic sequence of layers, implies that the three $\alpha_\nu$ 
parameters and the six $J_{\nu\nu'}$ parameters enter Eq.~(\ref{dvb}) in a linearly 
dependent way. We can then define
\begin{equation}
\widetilde{J}_{\nu\nu'} = J_{\nu\nu'} + \frac{\alpha_\nu + \alpha_{\nu'}}{2},
\end{equation}
in order to rewrite Eq.~(\ref{dvb}) as
\begin{equation}
A=\sum_{\langle\nu\nu'\rangle} N_{\nu\nu'} \widetilde{J}_{\nu\nu'} \;.
\label{dvbp}
\end{equation}
That is, we have eliminated
the $\alpha_\nu$ parameters; from now on, we consider our model to be
determined by the nine independent parameters
$\beta_C$, $\beta_S$, $\beta_B$, $\widetilde{J}_{CC}$, $\widetilde{J}_{SS}$,
$\widetilde{J}_{BB}$, $\widetilde{J}_{CS}$, $\widetilde{J}_{CB}$,
and $\widetilde{J}_{SB}$.

We obtain the values of the nine model parameters  $\{\beta_\nu, \widetilde{J}_{\nu\nu'} \}$ 
by fitting to the first-principles results for the six
two-component superlattices we considered. For each particular superlattice, the quadratic ($A$) and 
quatric ($B$) energy-decomposition coefficients, which are the linear combinations of 
$\{\beta_\nu, \widetilde{J}_{\nu\nu'} \}$ (see Table~\ref{AB_coeff} for explicit formulas), 
can be determined from first-principles superlattice polarization $p_\mathrm{min}$ and its 
ground-state energy $E(p_\mathrm{min})$ relative to the structure constrained to have zero 
polarization.\cite{fitting_comment1} These quantities are related to coefficients $A$ 
and $B$ as follows:

\begin{subequations}
\label{polar_fit}
\begin{eqnarray}
E(p_{min}) \equiv \Delta E = A p_\mathrm{min}^2 + B p_\mathrm{min}^4, \label{Eofp} \\
\left. \frac{dE(p)}{dp}\right|_{p=p_\mathrm{min}} \!\!\!\!\! = 0 \,\, 
\Longrightarrow \,\, A + 2 B p_\mathrm{min}^2 = 0. 
\label{dEofp}
\end{eqnarray}
\end{subequations}

\begingroup
\begin{table*}[ht!]
\tabcolsep=0.15cm
\renewcommand{\arraystretch}{1.3}

\caption{Fitting parameters used to predict polarization in CT/ST/BT superlattices. 
See text and Table~\ref{AB_coeff} for more details.}
\label{betaJ_coeff}
\begin{ruledtabular}
\begin{tabular}{ccccccccc}

$\beta_C$ & $\beta_S$ & $\beta_B$ & $\widetilde{J}_{CC}$ & $\widetilde{J}_{SS}$ & $\widetilde{J}_{BB}$ &
$\widetilde{J}_{CS}$ & $\widetilde{J}_{CB}$ & $\widetilde{J}_{SB}$ \\
 		       		       
\hline		       		       
		       		       
 \ph{-}0.584562 & \ph{-}0.648676 & \ph{-}0.833222 & -0.159671 & \ph{-}0.022111 & -0.219844 & -0.000834 & 
 -0.075701 & -0.079061 \\

\end{tabular}
\end{ruledtabular}
\end{table*}
\endgroup

The resulting parameters $\{\beta_\nu, \widetilde{J}_{\nu\nu'} \}$ are shown in 
Table~\ref{betaJ_coeff}.\cite{fitting_comment2} The \textit{ab initio} ground-state 
superlattice energies (relative to the corresponding nonpolar structures) and polarizations 
are shown in the fourth and fifth columns of Table~\ref{AB_coeff}. 
The values of fitted superlattice polarizations as well as the differences between them
and their \textit{ab initio} derived counterparts are shown in columns six and seven of 
the same table. The fitted polarization differences for the two-component superlattices 
are not presented, since they are, by construction, equal to the \textit{ab initio} ones. 
For the rest of the structures the model shows a remarkable agreement with first-principles 
results ($|\Delta p_\mathrm{min}| < 4\%$). For the three-component superlattices that possess 
inequivalent polarizations along [001] and [00$\bar{1}$] due to the breaking of inversion 
symmetry,\cite{sai2000,serge_APL} one could in principle compare with the larger polarization, 
the smaller one, or their average. We find empirically that the fit is best when compared with
the larger polarization, so we have chosen to present these values in the table.
The model performs poorly only for the strained bulk CT, whose first-principles value of 
polarization (computed in Ref.~\onlinecite{serge_APL}) represents an extreme limiting
case~\cite{CT_comment} and cannot be well reproduced by the model, which is fitted to the 
superlattice calculations.

The availability of such a convenient expression for computing polarization with nearly 
\textit{ab initio} precision allows us to predictively identify the arrangements 
of CT, ST and BT layers in a superlattice that would result in the largest possible 
polarization enhancement. While for short period superlattices ($N\leq 10$), this could 
be done by straightforward enumeration, the number of configurations increases rapidly
with $N$, necessitating a more sophisticated optimization procedure
for longer-period superlattices.  Here, we use a genetic algorithm,\cite{GAref}
in which a particular CT/ST/BT superlattice of a given
period $N$ is represented by a ``chromosome'' containing a sequence of
$C$, $S$ and $B$ ``genes''.  For example, a
(CT)$_2$(BT)$_1$(ST)$_2$(BT)$_1$ superlattice of period 6 is
encoded as a $CCBSSB$ chromosome.  The genetic algorithm also makes
it easy for us to impose constraints on the optimization, such as
limiting the thickness of individual layers or the average in-plane
lattice constant of the superlattice, as discussed further
below.

Our specific implementation of the genetic optimization algorithm is as follows. We create an 
initial population of $M$ chromosomes ($M$ is usually in between 2$N$ and 3$N$) by randomly assigning 
$C$, $S$ or $B$ values to each gene in each chromosome. The polarization of each chromosome is 
computed using Eq.~(7b), and the chromosome's ``fitness" 
is taken to be equal to the polarization.
The current generation of chromosomes is then replaced by the offspring-chromosome generation, 
created as follows.  First, the three chromosomes with the highest fitness (the so-called elite
chromosomes) are copied into the offspring generation without change to preserve the best
solutions from the previous generation.  Second, the remaining $M-3$ members of the next generation
are created by applying the following three-step procedure.
(i) Two ``parent'' chromosomes are selected from the current generation
by the so-called ``roulette wheel'' selection procedure,\cite{GAref_RWS}
which chooses a chromosome with a probability proportional to its fitness.
(ii) With probability 10\%, the offspring is taken to be identical to the parent with
better fitness.  The remaining 90\% of the time, a ``crossover'' procedure is applied.
We use a single-point crossover operator that randomly selects a single crossover 
point on the chromosome and copies the genes from one parent up to that point, and from
the other after that point.  (iii) Finally, the offspring
is subjected to a ``mutation'' operator, which changes the current value of each gene into one of
the two other available variants --- i.e., gene $S$, for example, could be changed to
either $C$ or $B$ --- with a low probability (in our case: 1\%). 
This entire selection and breeding process is continued for five 
hundred or more generations, after which the best available chromosomes are identified. For each 
set of parameters, i.e., the superlattice period and possible layer-sequencing restrictions, we 
perform five separate optimization runs to ensure convergence to a consistent solution.

\begingroup
\squeezetable
\begin{table}[b!]
\renewcommand{\arraystretch}{1.2}

\caption{Comparison between \textit{ab initio} polarizations $p_\mathrm{min}$ and fitted
polarizations $p_\mathrm{min}^{\mathrm{fit}}$ for a few polar short-period superlattices
identified by the genetic algorithm optimization procedure.}
\label{check}
\begin{ruledtabular}
\begin{tabular}{lcccc}
  System    &  $(N,k)$  & $|p_\mathrm{min}|$ (C/m$^2$) &
  $|p_\mathrm{min}^{\mathrm{fit}}|$ (C/m$^2$) & $|\Delta p_\mathrm{min}|$ (\%) \\
\hline
 (CT)$_1$(BT)$_3$          & (4,3) & 0.315 &	0.310 &	1.6 \\
 (ST)$_2$(BT)$_3$          & (5,3) & 0.279 &	0.275 & 1.2 \\
 (CT)$_2$(BT)$_4$          & (6,4) & 0.342 &	0.328 &	4.1 \\
 (CT)$_3$(ST)$_1$(BT)$_3$  & (7,3) & 0.313 &    0.305 & 2.6
 
\end{tabular}
\end{ruledtabular}
\end{table}
\endgroup

For any given $N$, if we impose no restrictions on the number of consecutively repeating 
layers of the same type, then the optimal configuration turns out to be pure CT or BT
(the fitted polarizations of bulk CT and BT are very close, see Table~\ref{AB_coeff}).
The former solution dominates in long period superlattices, while in shorter period 
ones ($N \le 10$) the latter solution is found more often. This happens because, as 
shown in Table~\ref{AB_coeff}, in thin superlattice layers BT has larger polarization 
than CT, which biases the optimization procedure towards BT.
However, as it is well known, neither of these configurations can be experimentally realized
because, when grown beyond a critical thickness, CT or BT relaxes to its natural
in-plane lattice constant and the strain-induced polarization enhancement is
lost. Thus, we constrain our optimization procedure 
so that only superlattices containing up to a given number $k$ of consecutive layers of the 
same type are allowed.

With this ``epitaxial growth'' constraint, the optimal superlattices 
that we find fall into two families depending on the relation between
$N$ and $k$. For even $N$ and $k \ge N/2$ or for odd $N$ and $k \ge (N-1)/2$,
the best solutions have the form of (XT)$_k$(YT)$_{N-k}$
or (XT)$_k$(YT)$_{N-k-1}$(ST)$_1$, where (X, Y) is (B, C) or (C, B).
On the other hand, optimal superlattices for smaller $k$ (relative to the same period $N$) 
contain a number of CT/ST/BT stripes and can be reduced to combinations of the 
best solutions of the same form as above but with smaller periods. For example, 
for $(N,k)$ = (12,4) we find three optimal superlattices with polarizations in the 
range of 0.32--0.33 C/m$^2$: (CT)$_4$(BT)$_4$(CT)$_2$(BT)$_2$, (CT)$_3$(BT)$_4$(CT)$_2$(BT)$_3$
and (CT)$_2$(BT)$_4$(CT)$_2$(BT)$_4$. Each of these superlattices splits into
two shorter ones with smaller $N$ and $k$. These are (8,4) (CT)$_4$(BT)$_4$ and (4,2) 
(CT)$_2$(BT)$_2$ for the first, (7,4) (CT)$_3$(BT)$_4$ and (5,3) (CT)$_2$(BT)$_3$ for the 
second, and two instances of (6,4) (CT)$_2$(BT)$_4$ for the third optimal superlattice, 
respectively. In what follows we restrict the discussion to solutions for large $k$ only,
assuming that in the opposite case optimal superlattices for any particular $N$ could
be constructed by merging together an appropriate number of the best large-$k$ solutions for
shorter periods.

We have carried out first-principles calculations for a few short-period
optimal superlattices to check that their \textit{ab initio} polarizations 
agree well with those predicted by the model.
We use a plane-wave based DFT-LDA method~\cite{pwscf} with ultrasoft 
pseudopotentials~\cite{USPP} for structural relaxation of the
superlattices and the Berry-phase method of the modern polarization 
theory~\cite{mpt} to compute their total polarization. The details of
the calculations are the same as in Ref.~\onlinecite{serge_APL}.
The results are presented in Table~\ref{check} and show that the good
agreement between \textit{ab initio} and fitted values of polarization in
short-period CT/ST/BT superlattices is preserved.

\begingroup
\begin{table}[ht!]
\renewcommand{\arraystretch}{1.2}

\caption{Short-period superlattices identified by the genetic algorithm 
optimization procedure as being the most polar and simultaneously having 
the lowest lattice constant mismatch $|\Delta a|$ with the substrate.}
\label{best}
\begin{ruledtabular}
\begin{tabular}{lcccc}
  System    &  $(N,k)$  & $|p_\mathrm{min}^{\mathrm{fit}}|$ (C/m$^2$) & $|\Delta a|$ (\%) \\
\hline

(CT)$_3$(BT)$_3$	    &   (6,3) &   0.327 &  0.34 \\
(CT)$_3$(ST)$_1$(BT)$_2$    &   (6,3) &   0.292 &  0.03 \\
(CT)$_3$(ST)$_1$(BT)$_3$    &   (7,3) &   0.305 &  0.29 \\
(CT)$_4$(BT)$_3$            &   (7,4) &   0.333 &  0.12 \\
(CT)$_4$(BT)$_4$            &   (8,4) &   0.337 &  0.34 \\
(CT)$_4$(ST)$_1$(BT)$_4$    &   (9,4) &   0.320 &  0.30 \\
(CT)$_5$(ST)$_1$(BT)$_4$    &  (10,5) &   0.324 &  0.15 \\
(CT)$_{10}$(ST)$_1$(BT)$_9$ & (20,10) &   0.346 &  0.24

\end{tabular}
\end{ruledtabular}
\end{table}
\endgroup

Another feature of the superlattice relevant to the feasibility
of its experimental realization is the mismatch between the 
equilibrium in-plane lattice constant of the 
superlattice (estimated by averaging over the unstrained lattice constants
of individual layers) and the lattice constant of the ST substrate.
The low substrate mismatch restriction tends to balance 
the number of CT ($a_\mathrm{CT} < a_\mathrm{ST}$) and BT ($a_\mathrm{BT} > a_\mathrm{ST}$) 
layers in the superlattice. With this additional screening step,
we find that the most polar CT/ST/BT superlattices
that emerge from the genetic optimization procedure 
have the following form: (CT)$_{N/2}$(BT)$_{N/2}$ for even $N$ and
(CT)$_{(N-1)/2}$(ST)$_1$(BT)$_{(N-1)/2}$ for odd $N$. It is worth
pointing out that adding one or two ST layers to CT and BT containing 
superlattices destroys their inversion symmetry without seriously
reducing polarization. The lack of the center of inversion makes 
the superlattice polarizations along [001] and [00$\bar{1}$] unequal,
which provides for even greater flexibility in fine-tuning of the polar 
properties of such structures.

In Table~\ref{best} we assemble a number of short-period CT/ST/BT superlattices
that were identified by the genetic algorithm optimization procedure as 
being the most polar superlattices with a lattice mismatch of less than
0.5\%, which should allow them to be grown coherently.\cite{ho_nyung}
The following first-principles lattice
constants were used for the substrate-mismatch analysis: $a_\mathrm{CT} = 3.813$ \AA\ 
(cubic), $a_\mathrm{ST} = 3.858$ \AA\ (cubic) and $a_\mathrm{BT} = 3.929$ \AA\ 
(tetragonal). On average, the polarizations of the superlattices presented in 
Table~\ref{best} are predicted to be 10--30\% higher than the computed polarizations 
of the previously investigated structures~\cite{serge_APL} shown in Table~\ref{AB_coeff}.

To conclude, we have used a first-principles-based one-dimensional chain model for polarization
in multicomponent perovskite-oxide ferroelectric superlattices combined with
a genetic algorithm optimization procedure to study the connection between the
polar properties of a superlattice and its layer sequence. We predict
specific layering arrangements that produce superlattices 
simultaneously possessing the highest possible polarization and a low in-plane
lattice-constant mismatch with the substrate. Our method could be applied to
superlattices containing individual components other than CT, ST and BT, or more
than three components, as long as the polarization profile across the superlattice 
remains sufficiently flat. Various additional restrictions on the arrangement of components 
could easily be added to the genetic algorithm optimization to design structures 
that are custom-tailored for specific applications.
Our predictions are for ideal
structures that are defect-free and fully switchable.
Since the remanent polarization of experimentally grown perovskite-oxide 
ferroelectric superlattices is substantially reduced due to structural defects and 
incomplete switching of ferroelectric domains, the computed values are expected
to be higher than those observed. Nevertheless, since our technique
does identify the most polar layer sequences (regardless of the absolute polarization) 
as well as quickly eliminating unfavorable arrangements, it can be used as a 
valuable tool to guide the experimental efforts in the quest for more efficient
nanoelectromechanical devices with tailored and/or substantially enhanced
properties.

The authors thank H. N. Lee for sharing his data and many valuable
discussions. This work was supported by the Center for
Piezoelectrics by Design (CPD) under ONR Grant N00014-01-1-0365.


\begin{thebibliography}{99}
\bibitem{fitting_comment1} For each superlattice we create
a corresponding non-polar structure with the ferroelectric displacement 
removed by ``unbuckling'' the AO and BO planes and moving the ``middle'' 
planes back to the center of each primitive cell.
\bibitem{fitting_comment2} We use both 
equations~(\ref{Eofp}) and~(\ref{dEofp}) for each two-component 
superlattice in the fit, which doubly overdetermins the set of 
$\{ \beta_\nu \}$.
\bibitem{CT_comment} We consider bulk CT under epitaxial tensile strain --- which
actually favors in-plane polarization --- but with ferroelectric displacement 
along the film-growth direction (symmetry restricted to space group P4mm) and 
all the zone boundary distortions that are present in its actual ground-state 
crystal structure suppressed. On the other hand, it is reasonable to assume that 
in a superlattice CT layers do develop substantial ferroelectric displacements in 
the film-growth direction by being polarized by the neighboring BT layers.
\end{thebibliography}
\end{document}